\documentclass[aps,twocolumn,showpacs]{revtex4}
\makeatletter

\newcommand{\Rmnum}[1]{\expandafter\@slowromancap\romannumeral #1@}
\makeatother
\usepackage{graphicx}
\usepackage{dcolumn}
\usepackage{bm}
\usepackage{CJK}
\usepackage{amsfonts}
\usepackage{psfrag}
\usepackage{wrapfig}
\usepackage{subfigure}
\usepackage{makeidx}
\usepackage{multirow}
\usepackage{epsf}
\usepackage{hyperref}
\usepackage{amsmath}
\usepackage{cases}
\usepackage[version=3]{mhchem}
\DeclareMathOperator{\sech}{sech}

\begin{document}
\title{Non-degenerate Bound State Solitons in Multi-component Bose-Einstein Condensates}
\author{Yan-Hong Qin$^{1,2}$}
\author{Li-Chen Zhao$^{1,2}$}\email{zhaolichen3@nwu.edu.cn}
\author{Liming Ling$^{3}$}\email{linglm@scut.edu.cn}
\address{$^{1}$School of Physics, Northwest University, Xi'an 710127, China}
\address{$^{2}$Shaanxi Key Laboratory for Theoretical Physics Frontiers, Xi'an 710127, China}
\address{$^{3}$School of Mathematics, South China University of Technology, Guangzhou 510640, China}
\date{\today}
\begin{abstract}
We investigate non-degenerate bound state solitons  systematically in multi-component Bose-Einstein condensates, through developing Darboux transformation method to derive exact soliton solutions analytically. In particular, we show that bright solitons with nodes correspond to the excited bound eigen-states in the self-induced effective quantum wells, in sharp contrast to the bright soliton and dark soliton reported before (which usually correspond to ground state and free eigen-state respectively). We further demonstrate that the bound state solitons with nodes are induced by incoherent interactions between solitons in different components. Moreover, we reveal that the interactions between these bound state solitons are usually inelastic, caused by the incoherent interactions between solitons in different components and the coherent interactions between solitons in same component. The bound state solitons can be used to discuss many different physical problems, such as beating dynamics, spin-orbital coupling effects, quantum fluctuations, and even quantum entanglement states.

\end{abstract}

\pacs{05.45.Yv, 02.30.Ik, 42.65.Tg}
\maketitle
\section{INTRODUCTION}
The multi-component coupled Bose-Einstein condensates (BECs) provide a good platform to study the dynamics of vector solitons \cite{sBEC}. Many different vector solitons have been obtained in the two-component coupled BEC systems, such as the bright-bright soliton\cite{BB1,BB2}, the bright-dark soliton \cite{BD}, the dark-antidark soliton \cite{D-antiD}, the dark-dark soliton \cite{DD1,DD2}, and the dark-bright soliton \cite{DB1,DB2}. The soliton states can be related with eigen-states in quantum well \cite{PCS,CPBzhao}. From the general properties of eigen-states in one-dimensional quantum wells, one can know that fundamental bright soliton corresponds to ground state and dark soliton is first-excited state in the effective quantum wells. Therefore, bright-bright soliton and dark-dark soliton are degenerate solitons (more than one component admits the same spatial mode), bright-dark soliton and dark-bright soliton are non-degenerate soliton states. The dark soliton state is a free state, and it admits a wide non-zero density background. This character is admitted by most of previous vector solitons in BECs systems\cite{BD,D-antiD,DD1,DD2,DB1,DB2}. We would like to look for non-degenerate bound state solitons (NDBSSs), for which all eigen-states are bound states. The bound state solitons can be used to investigate much more abundant beating or tunneling dynamics in multi-component BEC systems \cite{zhao2}, and discuss many other different physical problems, such as spin-orbital coupling effects \cite{SOC1,SOC2,SOC3}, quantum fluctuations \cite{QF1,QF2}, and even quantum entanglement states \cite{ET}.

In this paper, we obtain NDBSSs in BECs with attractive interactions, by performing Darboux transformation. Especially, we note that bright solitons with nodes correspond to the excited eigen-states in the effective quantum well. The incoherent interactions between solitons in different components can be seen as the mechanism of the bound state solitons. Furthermore, we investigate the interference properties of the NDBSSs. We show that the interference between solitons with nodes exhibits multi-periods, significantly differing with scalar solitons and bright-dark solitons. Moreover, our analysis reveal that the interactions between NDBSSs are inelastic in general, induced by the incoherent interactions between solitons in different components and the coherent interactions between solitons in same component. Double-hump and triple-hump solitons are demonstrated in two-component and three-component BECs respectively. These fascinating dynamics of non-degenerate solitons enrich the nonlinear dynamics in BECs system greatly, and the discussions on the mechanism of NDBBSs and their collision process further deepen our understanding on the vector solitons in BECs. Similar studies can be extended to more than three components cases, and more abundant bound state solitons are expected.

Our presentation of the above features will be structured as follows. In Sec. \Rmnum{2}, we introduce the theoretical model and present the NDBSS solutions. We further show that incoherent interactions between solitons in different components can be used to understand how come these bound state solitons.  In Sec.  \Rmnum{3}, we  reveal the collisions of NDBSSs are usually inelastic, due to the incoherent interactions and coherent interactions between these bound state solitons. In Sec.  \Rmnum{4}, we exhibit the NDBSS in three-component BECs systems. Finally, we summarize our results in Sec. \Rmnum{5}.

\section{Theoretical model and non-degenerate vector soliton solutions}
\begin{figure}[htbp]
\centering
{\includegraphics[width=86mm,height=26mm]{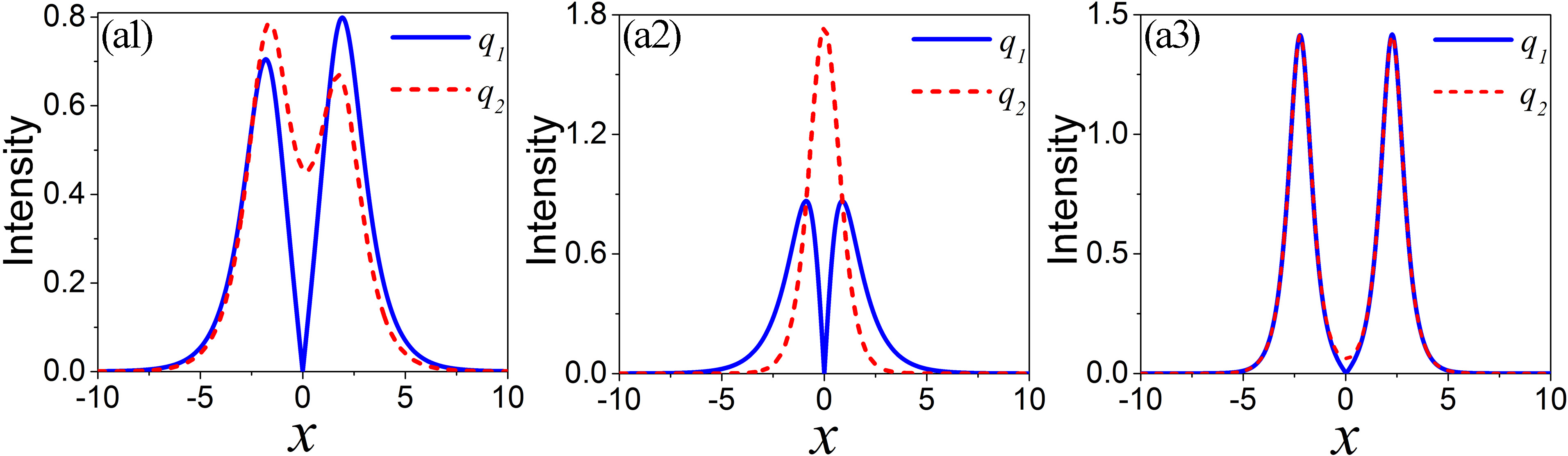}}
\caption{Three different density profiles for double-hump soliton in two-component coupled systems. (a1) for the asymmetric double-hump solitons in both components, (a2) for the symmetric single-hump-double-hump soliton, and (a3) for the approximate symmetric double-hump solitons in both components. Blue solid line and red dashed line correspond to first component and second component respectively. It is seen that solitons in $q_1$ component and $q_2$ component correspond to the first-excited state and ground state respectively in the effective quantum double-well.  The parameters for (a1) are $c_{11}=c_{22}=1,a_1=0,b_1=1,b_2=1.1,\delta=-2.8$, for (a2) are $c_{11}=c_{22}=\sqrt{3}/{3},a_1=0,b_1=1,b_2=2b_1,\delta=0$, and for (a3) are $c_{11}=c_{22}=1,a_1=0,b_1=2,b_2=2.001,\delta=-4.1$. }\label{Fig1}
\end{figure}

In the framework of mean-field theory, the dynamics of a two-component BEC in quasi-one dimension can be described well by the following dimensionless two-component coupled model \cite{zhaoliu1,MD1,MD2}:
\begin{equation}\label{CNLS1}
    \begin{split}
      {\rm i}q_{1,t}+q_{1,xx}+2(|q_1|^2+|q_2|^2)q_1&=0, \\
      {\rm i}q_{2,t}+q_{2,xx}+2(|q_1|^2+|q_2|^2)q_2&=0,
    \end{split}
\end{equation}
where $q_{1}$ and $q_{2}$ denote the two-component fields in the coupled BEC systems \cite{BEC}. The interactions between atoms are attractive for the above model, similar discussions can be done for repulsive interaction cases.  With the aid of Darboux transformation (DT) \cite{Mat,Dok,ling2,Lingdnls} or Hirota method \cite{Hirota,Lakshman}, many different vector solitons have been obtained in the two-component coupled BEC systems, such as the bright-bright soliton \cite{BB1,zhao1,ML1}, the bright-dark soliton \cite{BD}, the dark-dark soliton \cite{DD1,DD2}, and the dark-bright soliton \cite{DB1,DB2}. From the relations between soliton and eigen-states in quantum well \cite{PCS,CPBzhao}, we can know that bright-bright soliton and dark-dark soliton are degenerate solitons (more than one component admits the same spatial mode),  bright-dark soliton and dark-bright soliton are non-degenerate soliton states. However, the dark soliton state is a free state, and bright soliton usually admits no nodes \cite{BB1,zhao1,ML1}. This character is admitted by most of previous vector solitons in BECs systems. We would like to look for NDBSSs, for which all eigen-states are bound states, and bright solitons with nodes are present.

We develop DT method to derive the NDBSS. The twice DT with spectral parameters $\lambda_1=a_1+ib_1$ and $\lambda_2=a_1+ib_2$ generates one bound state soliton. The deriving  process is given in Appendix A in details, which is different from the processes for generating bright-bright soliton and bright-dark solitons \cite{Mat,Dok,ling2,Lingdnls}. The exact general double-hump soliton solution for \eqref{CNLS1} can be written as follows:
\begin{equation}\label{two-mode1}
\begin{split}
q_1(x,t)=-2ib_1c_{11}^{*}\frac{N_1}{M_1}e^{-i(a_1x+(a_1^2-b_1^2)t)}, \\
q_2(x,t)=-2ib_2c_{22}^{*}\frac{N_2}{M_1}e^{-i(a_1x+(a_1^2-b_2^2)t)}. \\
\end{split}
\end{equation}
with
\begin{align}
N_1&=\left(\frac{b_1-b_2}{b_1+b_2}+|c_{22}|^2e^{-2b_2(x+2a_1t)}\right)e^{-b_1(x+2a_1t)}, \nonumber \\
N_2&=\left(\frac{b_2-b_1}{b_1+b_2}+|c_{11}|^2e^{-2b_1(x+2a_1t)}\right)e^{-b_2(x+2a_1t)}, \nonumber \\
M_1&=|c_{11}|^2e^{-2b_1(x+2a_1t)}+|c_{22}|^2e^{-2b_2(x+2a_1t)} \nonumber \\
&+|c_{11}c_{22}|^2e^{-2(b_1+b_2)(x+2a_1t)}+\frac{(b_1-b_2)^2}{(b_1+b_2)^2}.\nonumber
\end{align}
where $a_1$, $b_1$, $b_2$ are real parameters, $c_{11}$ and $c_{22}$ are complex parameters. $-2a_1$ determines the velocity of soliton, and the parameters $b_1$, $b_2$, $|c_{11}|$, and $|c_{22}|$ determine the shape of soliton. In virtue of the expressions \eqref{two-mode1}, we see that the shape of soliton will be kept except the translation of position under the transform $|c_{11}|\to |c_{11}|e^{b_1\delta}$ and $|c_{22}|\to |c_{22}|e^{b_2\delta}$, $\delta$ is an arbitrary real constant. These parameters nontrivially contribute to the shape of soliton.  It is noted that the solitons in two components admit different modes. When $b_1<b_2$, soliton in $q_2$ component admits no node, and the one in $q_1$ component always has one node, and vice versa. From the general properties of bound states in one-dimensional potential \cite{QM}, we know that the eigen-state with one node corresponds to the first-excited state in the self-induced effective quantum well \cite{CPBzhao}. Therefore, the bound state solitons in the two components correspond to the ground state and the first-excited state in the effective quantum well respectively. Bound state soliton with  one node is in sharp contrast to the bright soliton and dark soliton reported before. In what follows, we discuss the profile types of the fundamental NDBSS.

The profiles of the bound state soliton can be mainly classified as three different types, asymmetric double-hump soliton (for which the two components both admit asymmetric double-hump), symmetric single-hump-double-hump soliton (for which one component admits single-hump soliton and the other component has a symmetric double-hump), and symmetric double-hump soliton (for which two components both admit symmetric double-hump). This classification is different from the one given in \cite{Laksh}, based on different aspects for soliton profiles.  The three different cases are shown in Fig.\ref{Fig1}(a1-a3), blue solid line and red dashed line corresponding to component $q_1$ and component $q_2$ respectively. (a1) depicts the intensity profile of asymmetric double-hump soliton in both components. The effective quantum well is $-2|q_1|^2-2|q_2|^2$, and it is a double-well form. The soliton in $q_1$ component and $q_2$ component correspond to the first-excited state and ground state respectively in the effective quantum double-well.  Particularly, we find that asymmetric double-hump soliton solution \eqref{two-mode1} can be reduced to symmetric form with parameters choice $c_{11}=c_{22}=\sqrt{3}/{3},b_2=2b_1,\delta=0$. For this case, we rewrite solutions as:
$q_1=\sqrt{3}b_1\sech[b_1(x+2a_1t)]\tanh[b_1(x+2a_1t)]e^{-i[a_1x+(a_1^2-b_1^2)t-\pi/2]}, q_2=\sqrt{3}b_1\sech[b_1(x+2a_1t)]^2e^{-i[a_1x+(a_1^2-4b_1^2)t+\pi/2]}$. The amplitude of component $q_1$ and $q_2$ are $\sqrt{3}b_1/2$ and $\sqrt{3}b_1$, respectively. One can see that a symmetric double-hump bright soliton present in component $q_1$, which corresponds to the first-exited bound state, while a single-hump ground state bright soliton emerge in component $q_2$.  This soliton can be seen as symmetric single-hump-double-hump soliton. As an example, we show it in Fig.\ref{Fig1}(a2).  Additionally, when $b_1$ and $b_2$ are very close to each other, the solutions \eqref{two-mode1} will show nearly symmetrical double-hump bright soliton in both components. A typical intensity profile is shown in Fig.\ref{Fig1}(a3). The two humps distribute symmetrically in each component for this case. Remarkably, it is clear that there is always one bright soliton with node in one component for all three cases. This character holds for these NDBSSs.

\begin{figure}[htbp]
\centering
{\includegraphics[width=85mm,height=70mm]{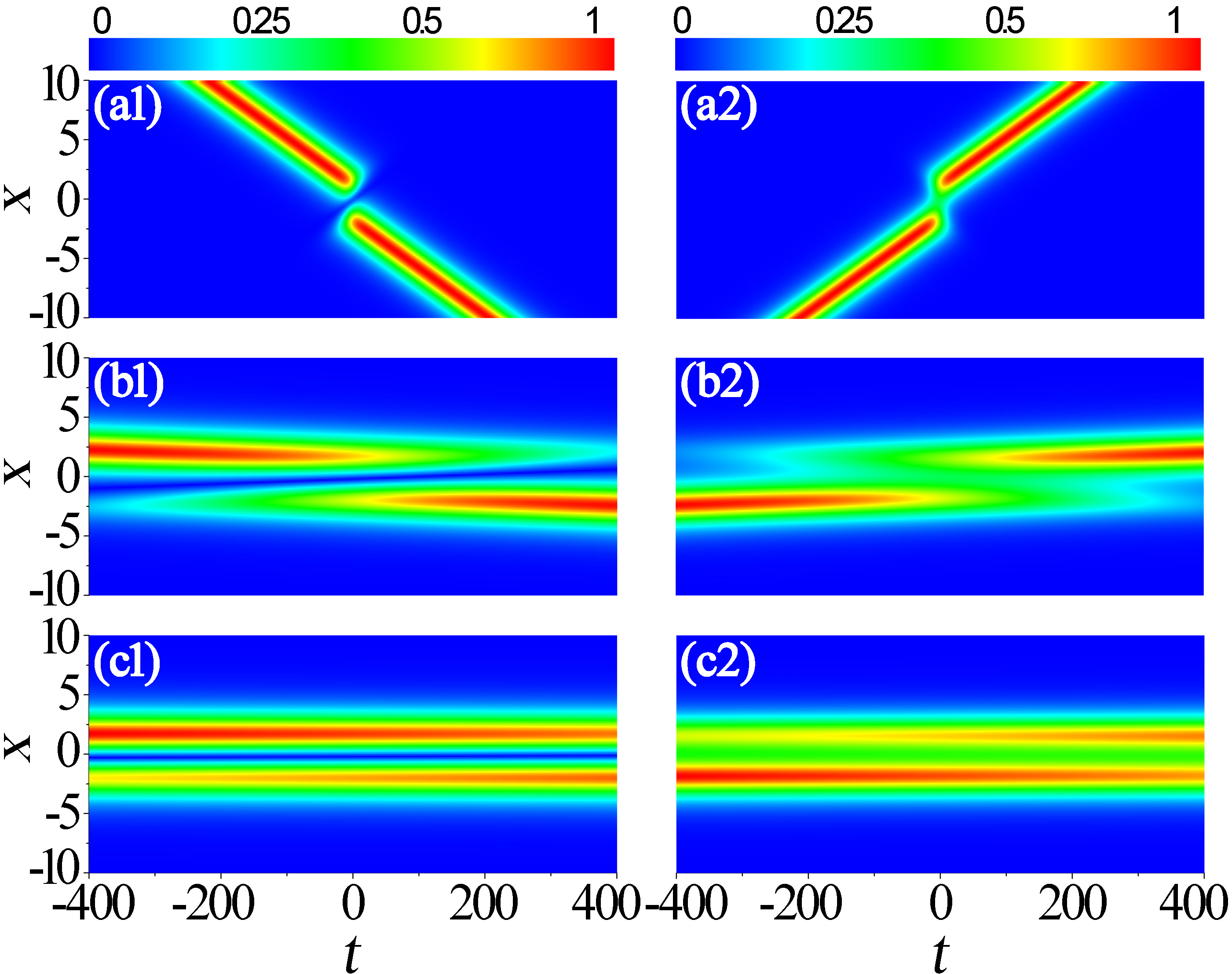}}
\caption{The incoherent interactions between solitons in different components with different relative velocities.  Form top to bottom the parameters are (a1) and (a2): $a_1=-a_2=0.02$, (b1) and (b2): $a_1=-a_2=0.001$, (c1) and (c2): $a_1=-a_2=0.0001$. Left panels show the density of first component, right panels the density of the second component. The evolution dynamics suggest that the non-degenerate bound state solitons is induced by incoherent superpositions of bright solitons in different components. The other parameters are $c_{11}=c_{22}=1,b_1=1,b_2=1.1,\delta=-2.8$ (same as Fig.\ref{Fig1} (a1)).}\label{Fig2}
\end{figure}

The soliton solution of the above Manakov model had been studied widely. However, the bright soliton with nodes are absent in the most of previously studies on vector solitons. Then, we would like to discuss how come the bound state soliton with nodes. Based on the deriving method for bound state soliton, we note that the bound state soliton is generated from two incoherent solitons with identical velocity.  The two incoherent solitons refer to the case for which each bright soliton just emerges in one component (different from the bright-bright solitons), and the two solitons in two components are at different locations (they are phase separated). For an example, one bright soliton moving to left in $q_1$ component, and there is no bright soliton in corresponding locations in the other component (see Fig. 2(a1)). This means the two solitons in two components just interact through the incoherent nonlinear interactions. The incoherent collision is different from the two solitons in one component for which the phases of solitons are coherent.  Therefore, we investigate the incoherent interactions between bright solitons in the two components, through varying the relative velocity. This can be done exactly by changing  the spectral parameters $\lambda_1=a_1+ib_1$ and $\lambda_2=a_2+ib_2$ in solution \eqref{eq:second transformation}. We change the velocity of soliton (i.e. $v_j=-2a_j$) in each component, and other parameters are fixed as Fig.\ref{Fig1}(a1). The relevant  dynamical processes of incoherent interactions between solitons are depicted in Fig.\ref{Fig2}, for which the relative velocity of solitons ($rv=v_1-v_2$) in two components corresponding to the $0.08$, $0.004$, and $0.0004$ respectively (see captions for detailed parameters setting). The two incoherent soliton character is shown clearly in Fig. 2(a). One can see that the relative velocity between solitons is becoming smaller, the incoherent collision between solitons in different components is becoming stronger. When the relative velocity of solitons decreases to zero, the general bright soliton in each components converts into double-hump soliton, such as solitons in Fig.\ref{Fig1}(a1). These dynamical processes indicate that NDBSSs are induced by the incoherent interactions between solitons in different components.

It should be mentioned that the similar soliton solutions have been found for a long time \cite{PCS}. Very recently, Hirota bilinear method was performed to derive similar non-degenerate vector solitons \cite{Laksh}. In this paper, we develop DT method to derive bound state solitons, which enables us to discuss the underlying mechanism for these bound state solitons. Moreover, the analysis uncover that bright soliton with one node corresponds to the first-excited state in the effective quantum well. The NDBSS involves the ground state and the first-excited state for the two-component cases. On the other hand, it was shown that many different static non-degenerate solitons were re-derived from the eigen-states in some certain quantum wells \cite{CPBzhao}. But multi-hump bright solitons are symbiotic with dark solitons, namely, the bound state and the free state are always coexist in the coupled systems. Those characters are different from the bound state solitons derived here. Especially,  those solutions generated from the eigen-states in quantum wells are stationary \cite{CPBzhao}, which are inconvenient to investigate the solitons collision analytically. We derive the more general NDBSS solutions systematically through developing DT method. Simultaneously, the collision processes between them can be investigated analytically in details.

\begin{figure}[htbp]
\centering
{\includegraphics[width=0.95\linewidth]{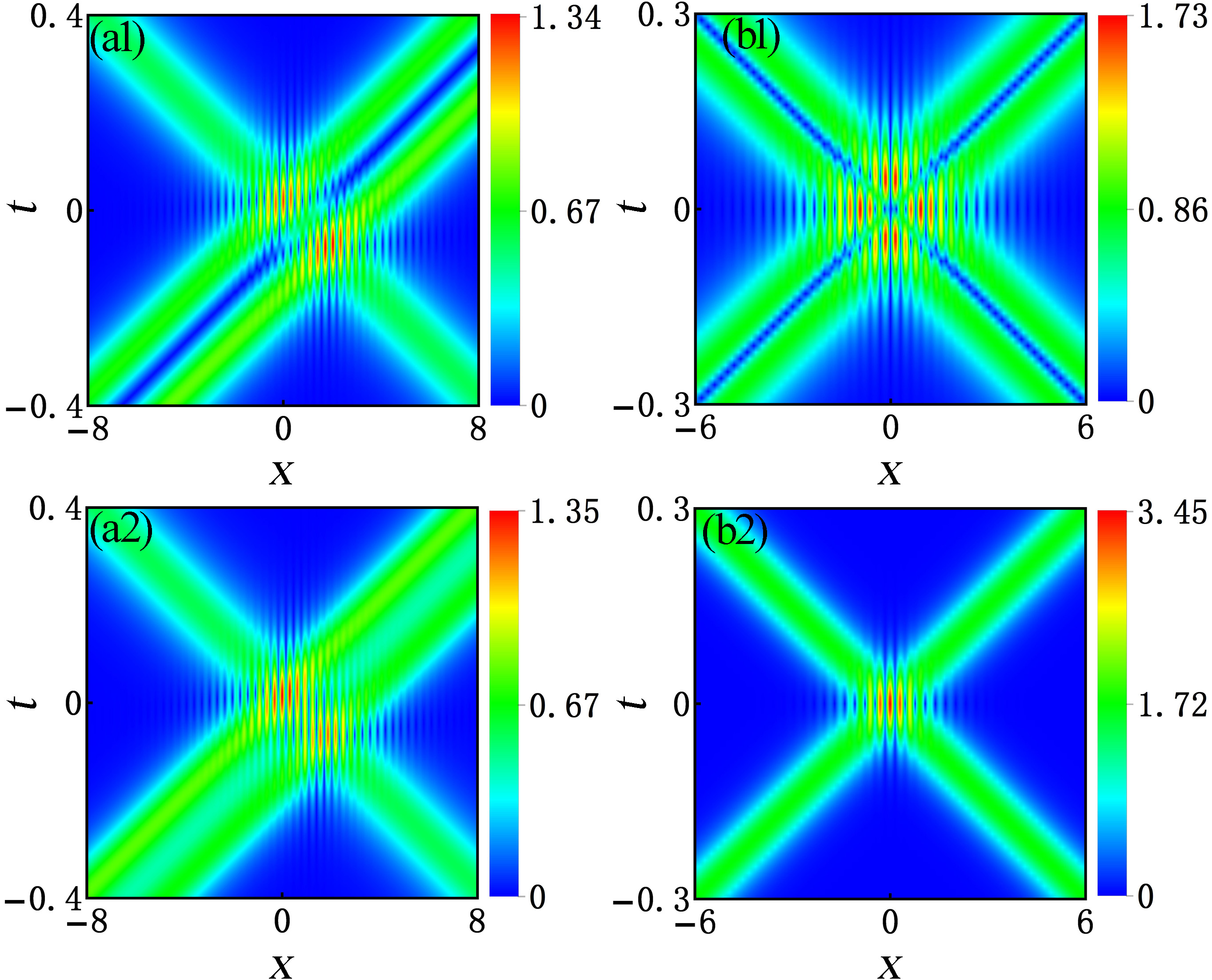}}
\caption{The collision dynamics between one non-degenerate bound state soliton and one degenerate bright soliton or one non-degenerate bound state soliton. (a1) and (a2): the interference behaviour between an asymmetric double-hump soliton (moving to right direction) and a degenerate bright soliton (moving to left).  Related parameters  are $c_{21}=c_{12}=0,c_{11}=c_{22}=c_{13}=c_{23}=1,a_1=-10,a_2=10,b_1=1,b_2=1.1,b_3=0.8,\delta_1=\delta_2=0$. (b1) and (b2): the interference patterns between two non-degenerate bound state solitons.  The parameters are $c_{11}=c_{22}=c_{13}=c_{24}=\sqrt{3}/{3},a_1=10,a_2=-10,b_1=1,b_2=2,b_3=1,b_4=2,\delta_1=\delta_2=0$. Top panels show the density of first component, bottom panels the density of the second component. The interference pattern is much abundant than the patterns between scalar bright solitons. It seems that the collisions are elastic. }\label{Fig3}
\end{figure}

\section{Collision between different non-degenerated solitons}
For simplicity and without losing generality, we investigate the interactions between a NDBSS and one degenerate bright soliton (one NDBSS) by performing third-fold DT (fourth-fold DT) (see the Appendix A for the detailed solving process). More complicated interaction cases between solitons can be investigated by performing N-fold DT in Appendix A \eqref{eq:n-fold-dt}.
Firstly, we investigate the collision between one NDBSS and a degenerate bright soliton, by performing third-fold DT with spectral parameters $\lambda_1=a_1+ib_1,\lambda_2=a_1+ib_2,$ and $\lambda_3=a_2+ib_3$. For this case, typical densities are depicted in the left panel of Fig.\ref{Fig3}, (a1) and (a2) correspond to component $q_1$ and component $q_2$ respectively. It is seen that the interference patterns between an asymmetric double-hump bright soliton and a single-hump bright soliton show in both components. But in component $q_1$ a first-excited bound eigen-state soliton interfere with a ground state soliton, while two ground state solitons collide with each other in component $q_2$. Detailed analysis indicate that the collisions between them are usually inelastic, and they can be elastic under some special initial conditions.

Secondly, we investigate the interaction between two NDBSSs by performing forth-fold DT with spectral parameters $\lambda_1=a_1+ib_1,\lambda_2=a_1+ib_2$ (generate one NDBSS), $\lambda_3=a_2+ib_3$, and $\lambda_4=a_2+ib_4$ (generate the other NDBSS).  We exhibit the dynamical evolution of them in the right panel of Fig.\ref{Fig3}, based on the two double-hump solitons solution \eqref{eq:fourth transformation}. (b1) and (b2) correspond to component $q_1$ and component $q_2$ respectively. As one can see in Fig.\ref{Fig3}(b1-b2), the collision of two identical symmetric double-hump solitons (two first-excited bound state solitons) in component $q_1$ and two identical single-hump solitons (two ground state solitons) in component $q_2$ all produce the interference patterns.   Moreover, we further explore the interference properties of them by asymptotic analysis technic (see the detailed solving process in Appendix A). Interestingly, we find that the interference of double-hump solitons presents multiperiodicity. The periodic functions are governed by the factors $\sin[(a_1-a_2)x+(a_1^2-a_2^2+b_3^2-b_1^2)t]$,  $\sin[(a_1-a_2)x+(a_1^2-a_2^2-b_2^2+b_4^2)t]$, and their corresponding cosine forms. This means that there are three periodic oscillation behaviours in the interference process of two double-hump bright solitons. The spatial period is $D=2\pi/(a_1-a_2)=4\pi/(v_1-v_2)$, and temporal periods are $T_1=2\pi/(a_1^2-a_2^2+b_3^2-b_1^2)$ and $T_2=2\pi/(a_1^2-a_2^2-b_2^2+b_4^2)$, in sharp contrast to interference pattern between bright solitons reported before \cite{NDzhao1}. This comes from the energy eigenvalues are more than two involving the interference process. In Fig.\ref{Fig3}(b), only the spatial interference pattern is visible, due to the parameters choice $a_1^2=a_2^2, b_1=b_3, b_2=b_4$ making two temporal periods all equal zero. For Fig.\ref{Fig3}(a1-a2), the parameters choice makes two temporal periods too small to be visible (see caption for more detail).

\begin{figure}[htbp]
\centering
\subfigure{\includegraphics[width=85mm,height=65mm]{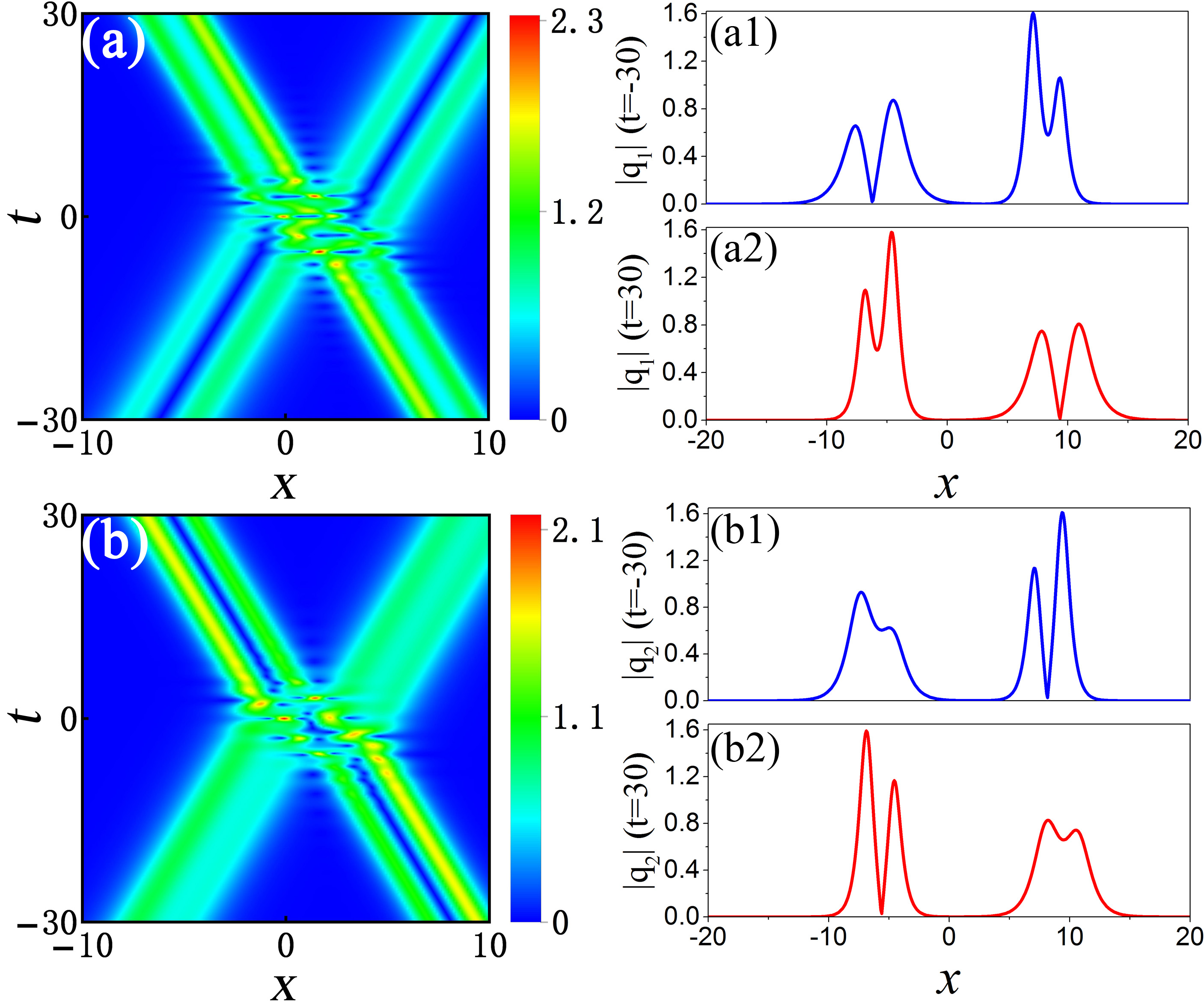}}
\caption{The inelastic collision between non-degenerate bound state solitons. Left panel: Dynamical evolution and coherent collision of non-degenerate bound state solitons. Right panel: Intensity plots for both components before (at t=-30, blue line) and after (at t=30, red line) the collision. It is seen that the profiles of double-hump solitons indeed change after collision. The analysis suggest that the collision between these bound state solitons are usually inelastic, due to the incoherent collision and coherent collision between them. The parameters are $c_{11}=c_{22}=c_{13}=c_{24}=1,a_1=-1/10,a_2=1/10,b_1=1,b_2=1.2,b_3=2,b_4=1.9,\delta_1=\delta_2=0.$}\label{Fig4}
\end{figure}

It seems that the collisions between two bound states solitons are elastic in Fig.\ref{Fig3}. Recent studies also suggested that  a fundamental double-hump soliton sustains its shape even after a collision with another similar soliton \cite{Laksh}. However, our studies suggest that the collision between bound state solitons is usually inelastic unless the parameters satisfy the sufficient condition \eqref{eq:elastic-cond}. In fact, the collisions in Fig.\ref{Fig3} are indeed inelastic, for which the soliton profiles change too slightly to be visible. A typical example for inelastic collision is shown in Fig.\ref{Fig4}, where two double-hump solitons collide with each other in both components. Left panel shows the density evolution, Fig.\ref{Fig4} (a) and (b) corresponding to component $q_1$ and $q_2$ respectively. Right panel depicts the intensity profile of two solitons before (at $t=-30$, blue line) and after ($t=30$, red line) the collision in both components. This figure makes it clear that the shape of double-hump solitons in each component possesses dramatic change after collision (see Fig.\ref{Fig4} (a1,a2) and (b1,b2)).  Then, what causes inelastic collisions of non-degenerate bound state solitons? As mentioned in the section II, the bound state solitons with nodes are induced by incoherent interactions between solitons in different components. We find that the incoherent interactions between solitons in different components and the coherent interplay between solitons in same component give rise to the inelastic collision of these bound state solitons. This can be seen clearly by investigating the interactions between solitons with different $\epsilon$ values (the spectral parameters are chosen as $\lambda_1=a_1+ib_1,\lambda_2=a_1+\epsilon_1+ib_2$, $\lambda_3=a_2+ib_3$, and $\lambda_4=a_2+\epsilon_2+ib_4$).

Experimental observations demonstrated that two-component solitons could be produced well based on well-developed density and phase modulation techniques \cite{DB1,DBST,DDMI}. Those experiments provide many hints that the above NDBSSs can be observed in two-component BECs. Very recently, three-component soliton states were further observed in a spinor BEC system \cite{TMB}. Motivated by these developments, we would like to extend our studies to three-component BECs for NDBSSs. Similar discussions can be extended to more than three components cases.

\begin{figure}[htbp]
\centering
\includegraphics[width=85mm,height=60mm]{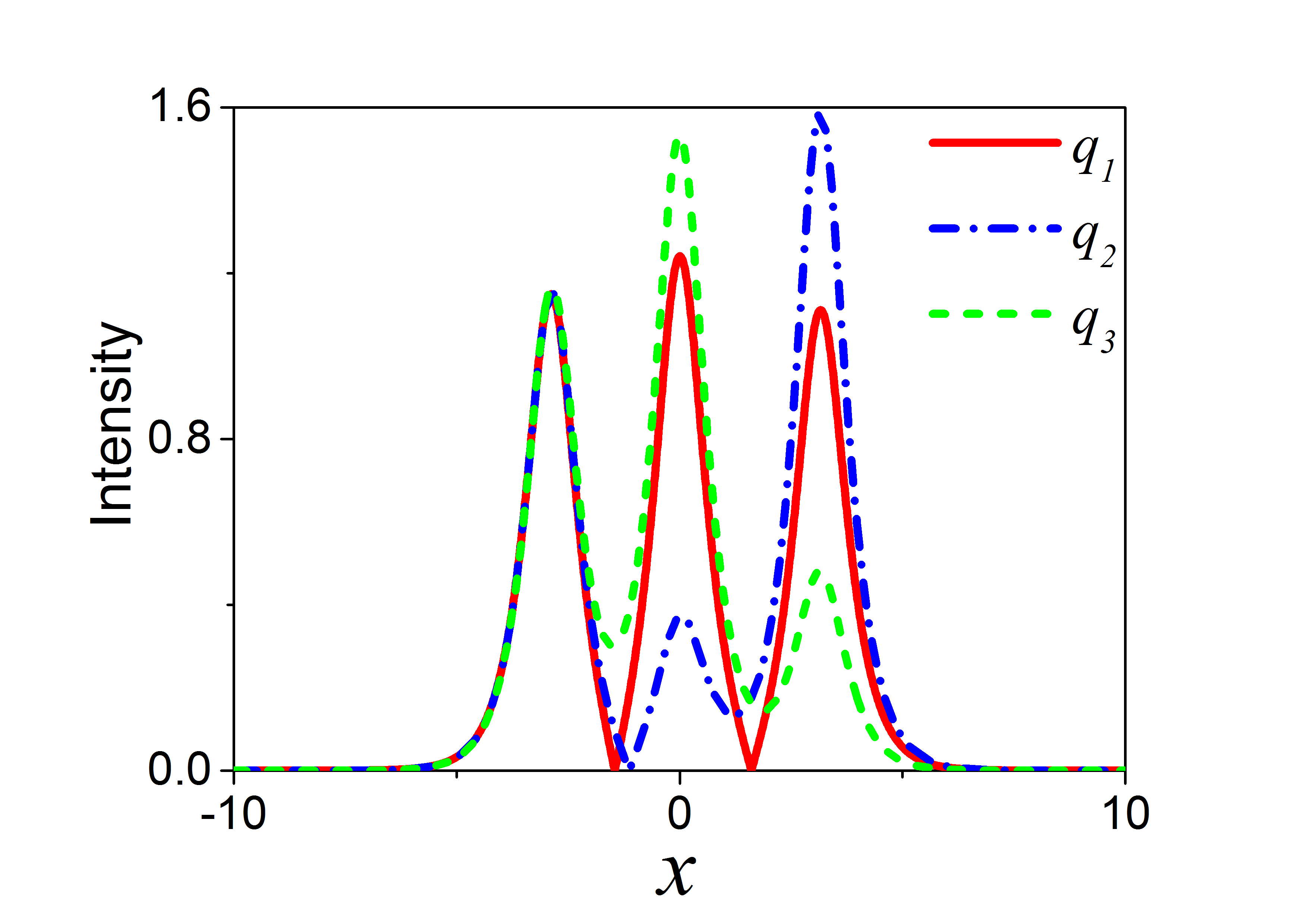}
\caption{Intensity profiles for triple-hump bright solitons in three-component coupled systems. The triple-hump bright soliton with no node in $q_3$ component is ground state (see green dashed line), the triple-hump bright soliton with one node in $q_2$ component is the first-excited state (see blue dotted-dashed line), and the triple-hump bright soliton with two nodes in $q_1$ component is the second-excited bound state soliton (see red solid line) in the effective quantum well. The parameters are $c_1=c_2=c_3=1,a_1=0,b_1=1.99,b_2=2,b_3=2.02,\delta=-5.2$. }\label{Fig5}
\end{figure}

\section{Triple-hump bright solitons in three-component condensates}
In this section, we consider the NDBSSs in three-component BECs system with attractive interactions. The dynamics can be described well by the following three-component coupled nonlinear equations in dimensionless form ($j=1,2,3$) \cite{sBEC}:
\begin{equation}\label{two-mode2}
    \begin{split}
      {\rm i}q_{j,t}+q_{j,xx}+2(|q_1|^2+|q_2|^2+|q_3|^2)q_j&=0
    \end{split}
\end{equation}
By direct performing the similar DT method with $\lambda_1=a_1+ib_1,\lambda_2=a_1+ib_2$, and $\lambda_3=a_1+ib_3$  as presented in Appendix A, the exact triple-hump bound state soliton solution of \eqref{two-mode2} can be written as follows (we have not presented the explicit solving process here for brevity):
\begin{equation}\label{eq:triple-hump}
    \begin{split}
      q_1(x,t)&=-2ib_1c_1e^{-i\alpha_1}(\frac{\chi_1}{\Xi_1}+\frac{2b_3\Delta_1\Delta_2}{(b_2+b_3)^2\Xi_2\Xi_1^2}e^{\beta_1}),\\
      q_2(x,t)&=-2ib_2c_2e^{-i\alpha_2}(\frac{\chi_2}{\Xi_1}+\frac{2b_3\Delta_1\Delta_3}{(b_2+b_3)^2\Xi_2\Xi_1^2}e^{\beta_2}),\\
      q_3(x,t)&=-2ib_3c_3\frac{\Delta_1e^{-i\alpha_3+\beta_3}}{(b_2+b_3)\Xi_2\Xi_1}.\\
    \end{split}
\end{equation}
where $\beta_j=-b_j(x+2a_1t),\alpha_j=a_1x+(a_1^2-b_j^2)t$. The explicit expressions of $\Xi_{1,2}$, $\chi_{1,2}$, $\Delta_{1,2,3}$ are given in the Appendix B. $a_1$, $b_1$, $b_2$, $b_3$ are real parameters, $c_1$, $c_2$, $c_3$ are complex parameters. $-2a_1$ determines the velocity of soliton, and the parameters $b_1$, $b_2$, $b_3$, $|c_1|$, $|c_2|$, $|c_3|$ govern the soliton profiles. In virtue of the expressions \eqref{eq:triple-hump}, we see that the shape of soliton will be kept except the translation of position under the transform $|c_1|\to |c_1|e^{b_1\delta}$, $|c_2|\to |c_2|e^{b_2\delta}$, $|c_3|\to |c_3|e^{b_3\delta}$, $\delta$ is an arbitrary real constant. These parameters nontrivially contribute to the shape of soliton.
The solution describes a general triple-hump soliton. A typical example of the intensity profile is displayed in Fig.\ref{Fig5}. It is seen that a triple-hump bright soliton exhibits in each component. The effective quantum well for this three-component case is $-2|q_1|^2-2|q_2|^2-2|q_3|^2$, and it is a triple-well form. Based on the the correspondence between solitons and eigen-states in quantum wells \cite{CPBzhao,QM}, one can know that the triple-hump bright soliton with no node in $q_3$ component is ground state (see green dashed line in Fig. 5), the triple-hump bright soliton with one node in $q_2$ component is the first-excited state (see blue dotted-dashed line in Fig. 5), and the triple-hump bright soliton with two nodes in $q_1$ component is the second-excited bound state soliton (see red solid line in Fig. 5) in the effective quantum well. Double-hump or single-hump solitons can be also obtained in the three-component case by choosing some proper parameters.  This suggests that more abundant NDBSSs can be found in more components coupled systems, since more components coupled BECs can induce deeper quantum wells. Similarly, the collision between triple-hump soliton and single-hump ground soliton can be investigated in three-component BECs by performing forth-fold DT.  The interaction between triple-hump soliton and double-hump soliton can be studied by performing fifth-fold DT. The interplay between two triple-hump bright solitons can be explored by performing sixth-fold DT. The inelastic collision of these bound state solitons can be also expected.

\section{conclusion}

In summary, we derive and investigate  double-hump and tripe-hump bound state solitons in multi-component BECs. The analysis indicates that bright solitons with nodes correspond to the excited bound eigen-states in the self-induced effective quantum wells. Particularly, we reveal that the incoherent interactions between solitons in different components is the generation mechanism of the bound state solitons. Furthermore, we demonstrate collisions of non-degenerate bound state solitons are inelastic in general case, which are induced by incoherent interactions and coherent interactions. Similar studies can be extended to more than two components cases, and more abundant bound state solitons are expected. These NDBSSs can be used to investigate much richer nonlinear dynamics and interactions in multi-component BEC systems, such as beating effects, tunneling dynamics, spin-orbital coupling effects, quantum fluctuations.

\section*{Acknowledgments}
This work is supported by National Natural Science Foundation of China (Contact No. 11775176), Basic Research Program of Natural Science of Shaanxi Province (Grant No. 2018KJXX-094), The Key Innovative Research Team of Quantum Many-Body Theory and Quantum Control in Shaanxi Province (Grant No. 2017KCT-12), and the Major Basic Research Program of Natural Science of Shaanxi Province (Grant No. 2017ZDJC-32).

\emph{Note added}:  Recently, we noticed  nondegenerate solitons were discussed in nonlinear optical fibers by the Hirota bilinear method \cite{Laksh}. In this paper, we perform Darboux transformation method to derive NDBSS solutions. Moreover, the discussions on the mechanism and the node properties could be helpful for our understanding on the NDBSS.

\section*{Appendix A:  The developed Darboux transformation method for deriving non-degenerate bound state soliton}
The two-component coupled nonlinear Schr\"{o}dinger equation \eqref{CNLS1} is the compatibility condition of the linear spectral problems \cite{ling2,zhaoliu1}:
\begin{equation}
    \begin{split}
     \Phi_x=U(x,t;\lambda)\Phi, \\
     \Phi_t=V(x,t;\lambda)\Phi, \\
    \end{split}
\end{equation}
where

\begin{equation}
    \begin{split}
U&=
\left(
  \begin{array}{ccc}
    -i\frac{2}{3}\lambda & q_1 & q_2\\
    -q_1^* & \frac{i}{3}\lambda & 0\\
    -q_2^* & 0 & \frac{i}{3}\lambda\\
    \end{array}
\right),\\
V&=U\lambda+
\left(
  \begin{array}{ccc}
   i|q_1|^2+i|q_2|^2 & iq_{1x} & iq_{2x} \\
   iq_{1x}^* & -i|q_1|^2 & -iq_2q_1^*\\
   iq_{2x}^* & -iq_2^*q_1 & -i|q_2|^2 \\
  \end{array}
\right)
    \end{split}
\end{equation}

The star denotes the complex conjugate. With the trivial seed solutions $q_1[0]=0,q_2[0]=0$ and spectral parameter $\lambda=\lambda_j=a_j+b_ji$ $(j=1,2,...,N)$, the vector eigenfunctions of the linear system Eqs.(5) can be writed as:
\begin{equation}\label{eq:eigenfunctions}
\Phi_j=\left(
  \begin{array}{c}
  \Phi_{1j}\\
  \Phi_{2j}\\
  \Phi_{3j}\\
\end{array}
\right)=
\left(
  \begin{array}{c}
   e^{-2\theta_j}\\
   c_{1j}e^{\theta_j}\\
   c_{2j}e^{\theta_j}\\
  \end{array}
\right),\,\,\,\, \theta_j=\frac{i}{3}\lambda_jx+\frac{i}{3}\lambda_j^2t,
\end{equation}
where $c_{1j},c_{2j}$ are the coefficients of eigenfunctions, and they are complex parameters. The fundamental one bright solitons can be obtained by the following Darboux transformation
\begin{equation}\label{eq:first transformation}
    \begin{split}
    \Phi[1]&=T[1]\Phi, T[1]=\mathbb{I}-\frac{\lambda_1-\lambda_1^*}{\lambda-\lambda_1^*}P[1]\\
     q_1[1]&=q_1[0]+(\lambda_1^*-\lambda_1)(P[1])_{12}\\
     q_2[1]&=q_2[0]+(\lambda_1^*-\lambda_1)(P[1])_{13}\\
    \end{split}
\end{equation}
$P[1]=\frac{\Phi_1\Phi_1^\dag}{\Phi_1^\dag\Phi_1}$, $\Phi_1$ is a special solution at $\lambda=\lambda_1$; a dagger denotes the matrix transpose and complex conjugate, and $(P[j])_{1j}$ represent the entry of matrix $P[j]$ in the first row and $j$ column. To obtain double-hump one soliton, we need to do the second step of transformation. We employ $\Phi_2$ which is mapped to $\Phi_2[1]=T[1]|_{\lambda=\lambda_2}\Phi_2$, one double-hump soliton solution can be obtained with spectral parameter $\lambda_2=a_1+ib_2$:
\begin{equation}\label{eq:second transformation}
    \begin{split}
    \Phi[2]&=T[2]\Phi[1],T[2]=\mathbb{I}-\frac{\lambda_2-\lambda_2^*}{\lambda-\lambda_2^*}P[2]\\
     q_1[2]&=q_1[1]+(\lambda_2^*-\lambda_2)(P[2])_{12}\\
     q_2[2]&=q_2[1]+(\lambda_2^*-\lambda_2)(P[2])_{13}\\
    \end{split}
\end{equation}
$P[2]=\frac{\Phi_2[1]\Phi_2[1]^\dag}{\Phi_2[1]^\dag\Phi_2[1]}$. For this case, we choose the coefficients of eigenfunctions $\Phi_{1,2}$ \eqref{eq:eigenfunctions} as the following way: (i) $c_{21}=c_{12}=0$, $c_{11},c_{22}$ are nonzero complex parameters, or (ii) $c_{11}=c_{22}=0$ and $c_{21},c_{12}$ are nonzero complex parameters. The corresponding simplified solution has been present in \eqref{two-mode1}. Examples of the relevant intensity profiles have been exhibit in Fig.\ref{Fig1}.

To study the interaction between non-degenerate solitons, it needs to do multiple step transition. For example, by performing the third-step of transition, we employ $\Phi_3$ which is mapped to $\Phi_3[2]=(T[2]\Phi_3[1])|_{\lambda=\lambda_3}$ with $\Phi_3[1]=(T[1]\Phi_3)|_{\lambda=\lambda_3}$, then the collision between a double-hump soliton and a single-hump soliton can be obtained with spectral parameter $\lambda_3=a_2+ib_3$:
\begin{equation}\label{eq:third transformation}
    \begin{split}
    \Phi[3]&=T[3]\Phi[2],T[3]=\mathbb{I}-\frac{\lambda_3-\lambda_3^*}{\lambda-\lambda_3^*}P[3]\\
     q_1[3]&=q_1[2]+(\lambda_3^*-\lambda_3)(P[3])_{12}\\
     q_2[3]&=q_2[2]+(\lambda_3^*-\lambda_3)(P[3])_{13}\\
    \end{split}
\end{equation}
$P[3]=\frac{\Phi_3[2]\Phi_3[2]^\dag}{\Phi_3[2]^\dag\Phi_3[2]}$. For this case, we choose the coefficients of eigenfunctions $\Phi_3$ \eqref{eq:eigenfunctions} as the following way: (i) $c_{13},c_{23}$ are nonzero complex parameters, or (ii) $c_{13}=0$ or (iii) $c_{23}=0$, and the coefficients of eigenfunctions $\Phi_{1,2}$ are same as the \eqref{eq:second transformation}. Typical example for this case has been shown in Fig.\ref{Fig3} (a1) and (a2).

Naturally, by performing the fourth-step transformation, one can investigate the interaction between two double-hump solitons. We employ $\Phi_4$ which is mapped to $\Phi_4[3]=(T[3]\Phi_4[2])|_{\lambda=\lambda_4}=(T[3]T[2]T[1]\Phi_4)|_{\lambda=\lambda_4}$, then two double-hump solitons solutions can be obtained as follows with spectral parameter $\lambda_4=a_2+ib_4$ :
\begin{equation}\label{eq:fourth transformation}
    \begin{split}
     q_1[4]&=q_1[3]+(\lambda_4^*-\lambda_4)(P[4])_{12}\\
     q_2[4]&=q_2[3]+(\lambda_4^*-\lambda_4)(P[4])_{13}\\
    \end{split}
\end{equation}
$P[4]=\frac{\Phi_4[3]\Phi_4[3]^\dag}{\Phi_4[3]^\dag\Phi_4[3]}$. For this case, the coefficients of vector eigenfunctions $\Phi_{3,4}$ are analogous to $\Phi_{1,2}$, i.e., (i) $c_{23}=c_{14}=0$, $c_{13},c_{24}$ are nonzero complex parameters, or (ii) $c_{13}=c_{24}=0$ and $c_{23},c_{14}$ are nonzero complex parameters. One typical case has been shown Fig.\ref{Fig3} (b1) and (b2).

In general, the $N$-fold Darboux matrix can be constructed as the following form:
\begin{equation}\label{eq:n-fold-dt}
\begin{split}
\mathbf{T}_N&=\mathbb{I}-\mathbf{X}_N\mathbf{M}_N^{-1}(\lambda\mathbb{I}-\mathbf{D}_N)^{-1}\mathbf{X}_N^{\dag},\\
\mathbf{X}_N&=\left[\Phi_1,\Phi_2,\cdots, \Phi_N\right],\\
\mathbf{D}_N&={\rm diag}\left(\lambda_1^*,\lambda_2^*,\cdots,\lambda_N^*\right), \\
\mathbf{M}_N&=\left(\frac{\Phi_i^{\dag}\Phi_j}{\lambda_j-\lambda_i^*}\right)_{1\leq i,j\leq N},
\end{split}
\end{equation}
and the B\"acklund transformation between old potential functions and new ones are
\begin{equation}\label{eq:back-n-fold}
\begin{split}
 q_1[N]&=q_1+\frac{\det(\mathbf{M}_{N,1})}{\det(\mathbf{M}_N)} \\
 q_2[N]&=q_2+\frac{\det(\mathbf{M}_{N,2})}{\det(\mathbf{M}_N)} \\
\end{split}
\end{equation}
where
\[
\mathbf{M}_{N,1}=\begin{bmatrix}
\mathbf{M}_N&\mathbf{X}_{N,2}^{\dag} \\[5pt]
\mathbf{X}_{N,1}&0\\
\end{bmatrix},\,\,\,\,\mathbf{M}_{N,2}=\begin{bmatrix}
\mathbf{M}_N&\mathbf{X}_{N,3}^{\dag} \\[5pt]
\mathbf{X}_{N,1}&0\\
\end{bmatrix}
\]
$\mathbf{X}_{N,i}$ represents the $i$-th row of matrix $\mathbf{X}_N$.

For the two double-hump soliton, we choose the parameters as the following way: $\lambda_1=a_1+b_1 i$, $\lambda_2=a_1+b_2 i$, $c_{21}=0$, $c_{12}=0$, $c_{11}$, $c_{22}$ are non zero complex parameters, which determine the first double-hump soliton; and $\lambda_3=a_2+b_3 i$, $\lambda_4=a_2+b_4 i$, $c_{23}=0$, $c_{14}=0$, $c_{13}$, $c_{24}$ are non zero complex parameters, which determine the second one. The oscillator for the two solitons is governed by the factors $\sin[(a_1-a_2)x+(a_1^2-a_2^2+b_3^2-b_1^2)t]$, $\cos[(a_1-a_2)x+(a_1^2-a_2^2+b_3^2-b_1^2)t]$ and $\sin[(a_1-a_2)x+(a_1^2-a_2^2-b_2^2+b_4^2)t]$, $\cos[(a_1-a_2)x+(a_1^2-a_2^2-b_2^2+b_4^2)t]$. The velocity of solitons is controlled by $x+2a_j t={\rm const}$, $j=1,2,$ respectively, i.e. the velocity of soliton equals to $-2a_j$. Assume that $b_j>0$, $j=1,2,3,4$, and $a_1>a_2$, fixed the parameters of the first double-hump soliton $x+2a_1t={\rm const}$, then $x+2a_2t=x+2a_1t+2(a_2-a_1)t$. If $t\to +\infty$, we have $x+2a_2t\to -\infty$. Then we see that
\[
\Phi_3\parallel \left(
  \begin{array}{c}
   e^{-i\lambda_jx-i\lambda_j^2t}\\
   c_{13}\\
   0\\
  \end{array}
\right)\to \left(
  \begin{array}{c}
   0 \\
   c_{13}\\
   0\\
  \end{array}
\right)
\]
and
\[
\Phi_4\parallel \left(
  \begin{array}{c}
   e^{-i\lambda_jx-i\lambda_j^2t}\\
   0\\
   c_{24}\\
  \end{array}
\right)\to \left(
  \begin{array}{c}
   0 \\
   0\\
   c_{24}\\
  \end{array}
\right)
\]
Since the order of iteration for the Darboux transformation can be exchanged, we rewrite
\[
T[1]=\mathbb{I}-\frac{\lambda_3-\lambda_3^*}{\lambda-\lambda_3^*}
\frac{\Phi_3\Phi_3^{\dag}}{\Phi_3^{\dag}\Phi_3}\to {\rm diag}\left(1,\frac{\lambda-\lambda_3}{\lambda-\lambda_3^*},1\right)
\]
along the line of $x-a_1 t={\rm const}$ as $t\to \infty$. It follows that
\[
\Phi_4[1]\parallel T[1]|_{\lambda=\lambda_4}\left(
  \begin{array}{c}
   0 \\
   0\\
   c_{24}\\
  \end{array}
\right)\to \left(
  \begin{array}{c}
   0 \\
   0\\
   c_{24}\\
  \end{array}
\right)
\]
which deduces that
\[
T[2]\to {\rm diag}\left(1,1,\frac{\lambda-\lambda_4}{\lambda-\lambda_4^*}\right).
\]
Combining the first and second Darboux matrix, we obtain \[
T[2]T[1]\to {\rm diag}\left(1,\frac{\lambda-\lambda_3}{\lambda-\lambda_3^*},\frac{\lambda-\lambda_4}{\lambda-\lambda_4^*}\right),
\]
which yields that
\begin{multline}
\Phi_1[2]=T[2]T[1]|_{\lambda=\lambda_1}\Phi_1\to \Phi_1^{[+]}=
\left(
  \begin{array}{c}
   e^{-2\theta_1}\\
   c_{11}^{[+]}e^{\theta_1}\\
   0 \\
  \end{array}
\right),\\
\Phi_2[2]=T[2]T[1]|_{\lambda=\lambda_2}\Phi_2\to \Phi_2^{[+]}=
\left(
  \begin{array}{c}
   e^{-2\theta_2}\\
   0 \\
   c_{22}^{[+]}e^{\theta_2}\\
  \end{array}
\right)
\end{multline}
where $c_{11}^{[+]}=\frac{\lambda_1-\lambda_3}{\lambda_1-\lambda_3^*} c_{11}$, $c_{22}^{[+]}=\frac{\lambda_2-\lambda_4}{\lambda_2-\lambda_4^*} c_{22}.$

Thus, when $t\to +\infty$, the Darboux matrix tends to:
\begin{multline}\label{eq:analysis-dt}
\mathbf{T}_4\to
\left(\mathbb{I}-\mathbf{X}_{2}^{[+]} (\mathbf{M}_{2}^{[+]})^{-1}(\lambda\mathbb{I}-\mathbf{D}_{2})^{-1}(\mathbf{X}_{2}^{[+]})^{\dag}\right)
\\ \times {\rm diag}\left(1,\frac{\lambda-\lambda_3}{\lambda-\lambda_3^*},\frac{\lambda-\lambda_4}{\lambda-\lambda_4^*}\right) \\
\end{multline}
where
\[
\mathbf{X}_{2}^{[+]}=\left[\Phi_1^{[+]},\Phi_2^{[+]}\right],\,\,\, \mathbf{D}_{2}={\rm diag}\left(\lambda_1^*,\lambda_2^*\right)
\]
and
\[
\mathbf{M}_{2}^{[+]}=
\begin{bmatrix}
\frac{1+\left|c_{11}^{[+]}\right|^2e^{6{\rm Re}(\theta_1)}}{\lambda_1-\lambda_1^*} & \frac{1}{\lambda_2-\lambda_1^*}\\[8pt]
\frac{1}{\lambda_1-\lambda_2^*} &\frac{1+\left|c_{22}^{[+]}\right|^2e^{6{\rm Re}(\theta_2)}}{\lambda_2-\lambda_2^*} \\
\end{bmatrix}.
\]
From the Darboux matrix \eqref{eq:analysis-dt}, we obtain that the double-hump soliton approaches to
\begin{equation}
\begin{split}
q_1[4]\to&q_1(x,t;a_1,b_1,b_2,c_{11}^{[+]},c_{22}^{[+]}), \\
q_2[4]\to&q_2(x,t;a_1,b_1,b_2,c_{11}^{[+]},c_{22}^{[+]}) \\
\end{split}
\end{equation}
as $t\to\infty$ along the line $x+2a_1t={\rm const}$, where $q_1$, $q_2$ are given in equations \eqref{two-mode1}.
In a similar manner, as $t\to-\infty$, we have
\begin{equation}
\begin{split}
q_1[4]\to&q_1(x,t;a_1,b_1,b_2,c_{11}^{[-]},c_{22}^{[-]}), \\
q_2[4]\to&q_2(x,t;a_1,b_1,b_2,c_{11}^{[-]},c_{22}^{[-]}) \\
\end{split}
\end{equation}
as $t\to-\infty$ along the line $x+2a_1t={\rm const}$, where
$c_{11}^{[-]}=\frac{\lambda_1-\lambda_3^*}{\lambda_1-\lambda_3}\frac{\lambda_1-\lambda_4^*}{\lambda_1-\lambda_4} c_{11}$, $c_{22}^{[-]}=\frac{\lambda_2-\lambda_3^*}{\lambda_2-\lambda_3}\frac{\lambda_2-\lambda_4^*}{\lambda_2-\lambda_4} c_{22}.$

Now we consider the asymptotic behavior of the second soliton. Fixed $x+2a_2t={\rm const}$, as $t\to \pm\infty$, we have the asymptotic expression
\begin{equation}
\begin{split}
q_1[4]\to&q_1(x,t;a_2,b_3,b_4,c_{13}^{[\pm]},c_{24}^{[\pm]}),\\
q_2[4]\to&q_2(x,t;a_2,b_3,b_4,c_{13}^{[\pm]},c_{24}^{[\pm]})\\
\end{split}
\end{equation}
where
$c_{13}^{[+]}=\frac{\lambda_3-\lambda_1^*}{\lambda_3-\lambda_1}\frac{\lambda_3-\lambda_2^*}{\lambda_3-\lambda_2} c_{13}$, $c_{24}^{[+]}=\frac{\lambda_4-\lambda_1^*}{\lambda_4-\lambda_1}\frac{\lambda_4-\lambda_2^*}{\lambda_4-\lambda_2} c_{24}$ and $c_{13}^{[-]}=\frac{\lambda_3-\lambda_1}{\lambda_3-\lambda_1^*} c_{13}$, $c_{24}^{[-]}=\frac{\lambda_4-\lambda_2}{\lambda_4-\lambda_2^*} c_{24}.$

In general case, the interaction between two hump soliton is still inelastic. But under the special case,
\begin{equation}\label{eq:elastic-cond}
\begin{split}
|c_{11}^{[+]}|&=|c_{11}^{[-]}|e^{b_1\delta_1},\,\,\,\,\, |c_{22}^{[+]}|=|c_{22}^{[-]}|e^{b_2\delta_1},\\
|c_{13}^{[+]}|&=|c_{13}^{[-]}|e^{b_3\delta_2},\,\,\,\,\, |c_{24}^{[+]}|=|c_{24}^{[-]}|e^{b_4\delta_2},
\end{split}
\end{equation}
the interaction is elastic. In other words, this is the sufficient condition \eqref{eq:elastic-cond} of elastic interaction of two-hump soliton.

\section*{Appendix B: The expressions of $\Xi_{1,2}$, $\chi_{1,2}$, $\Delta_{1,2,3}$ }

The general one triple-hump soliton solution in three-component NLSE is expressed as \eqref{eq:triple-hump}, where $\Xi_{1,2}$, $\chi_{1,2}$, $\Delta_{1,2,3}$ are written as

\begin{equation}\label{two-mode4}
\begin{split}
\Xi_1&\!=\!\frac{(b_1-b_2)^2}{(b_1+b_2)^2}\!+\!|c_1|^2e^{2\beta_1}\!+\!|c_{2}|^2e^{2\beta_2}\!+\!|c_1|^2|c_2|^2e^{2(\beta_1+2\beta_2)},\\
\Xi_2&=\frac{\Delta_1^2}{(b_2+b_3)^2\Xi_1^2}\!+\!\frac{4b_1^2|c_1|^2\Delta_2^2}{(b_2+b_3)^2\Xi_1^2}e^{2\beta_1}\!+\!\frac{4b_2^2|c_2|^2\Delta_3^2}{(b_2+b_3)^2\Xi_1^2}e^{2\beta_2}\\
&+\!|c_3|^2e^{2\beta_3},\\
\Delta_{1}&\!=\!(b_2-b_3)\left[\frac{(b_1-b_2)^2(b_1-b_3)}{(b_1+b_2)^2(b_1+b_3)}-|c_1|^2e^{2\beta_1}\!\right]\\
&\!+\!(b_2+b_3)\left[\frac{b_3-b_1}{b_3+b_1}+|c_1|^2e^{2\beta_1}\right]|c_2|^2e^{2\beta_2},\\
\Delta_{2}&\!=\!\frac{(b_1-b_2)(b_2-b_3)}{(b_1+b_2)(b_1+b_3)}-\frac{b_2+b_3}{b_1+b_3}|c_2|^2e^{2\beta_2},\\
\Delta_{3}&\!=\!\frac{(b_2^2-b_1^2)(b_1-b_3)}{(b_1+b_2)^2(b_1+b_3)}-|c_1|^2e^{2\beta_1},\\
\chi_{1}&=\left[\frac{b_1^2-b_2^2}{(b_1+b_2)^2}+|c_2|^2e^{2\beta_2}\right]e^{\beta_1},\\
\chi_{2}&=\left[\frac{b_2^2-b_1^2}{(b_1+b_2)^2}+|c_1|^2e^{2\beta_1}\right]e^{\beta_2}.\\\nonumber
\end{split}
\end{equation}

\end{document}